\documentclass[sigconf,natbib=true]{acmart}

\usepackage{tikz}
\usepackage{array,multirow,graphicx}
\usepackage{adjustbox}
\usepackage{amsmath}
\usepackage{bm}
\usepackage{colortbl}
\usepackage{enumitem}
\usepackage{tabu}
\usepackage{balance}
\usepackage{footnote}
\usepackage{pifont}
\usepackage{algorithm}
\usepackage{algpseudocode}
\usepackage{subfig}
\usepackage{tikz}

\setlist[itemize]{topsep=1pt}
\usepackage[font=small,skip=0pt]{caption}
\AtBeginDocument{%
  \providecommand\BibTeX{{%
    \normalfont B\kern-0.5em{\scshape i\kern-0.25em b}\kern-0.8em\TeX}}}

\copyrightyear{2023} 
\acmYear{2023} 
\setcopyright{acmlicensed}\acmConference[SIGIR-AP '23]{Proceedings of the 1st International ACM SIGIR Conference on Information Retrieval in the Asia Pacific}{November 26--29, 2023}{Beijing, China}
\acmBooktitle{Proceedings of the 1st International ACM SIGIR Conference on Information Retrieval in the Asia Pacific (SIGIR-AP'23, November 26--29, 2023, Beijing, China}
\acmPrice{15.00}
\acmISBN{xxxx}
\acmDOI{doi.org/xxxxxx}

\begin{document}



\title{Annotating Data for Fine-Tuning a Neural Ranker? Current Active Learning Strategies are not Better than Random Selection}


\author{Sophia Althammer}
\email{sophia.althammer@tuwien.ac.at}
\affiliation{%
  \institution{TU Wien, Austria}
  \country{}
}

\author{Guido Zuccon}
\email{g.zuccon@uq.edu.au}
\affiliation{%
  \institution{University of Queensland, Australia}
  \country{}
}

\author{Sebastian Hofst{\"a}tter}
\email{sebastian.hofstaetter@tuwien.ac.at}
\affiliation{%
  \institution{Cohere, Austria}
  \country{}
}

\author{Suzan Verberne}
\email{s.verberne@liacs.leidenuniv.nl}
\affiliation{%
  \institution{Leiden University, Netherlands}
  \country{}
}

\author{Allan Hanbury}
\email{allan.hanbury@tuwien.ac.at}
\affiliation{%
 \institution{TU Wien, Austria}
 \country{}
 }

\renewcommand{\shortauthors}{Althammer et al.}

\begin{abstract}
Search methods based on Pretrained Language Models (PLM) have demonstrated great effectiveness gains compared to statistical and early neural ranking models. However, fine-tuning PLM-based rankers requires a great amount of annotated training data. Annotating data involves a large manual effort and thus is expensive, especially in domain specific tasks.
In this paper we investigate fine-tuning PLM-based rankers under limited training data and budget. 
We investigate two scenarios: fine-tuning a ranker from scratch, and domain adaptation starting with a ranker already fine-tuned on general data, and continuing fine-tuning on a target dataset.

We observe a great variability in effectiveness when fine-tuning on different randomly selected subsets of training data. This suggests that it is possible to  achieve effectiveness gains by actively selecting a subset of the training data that has the most positive effect on the rankers. This way, it would be possible to fine-tune effective PLM rankers at a reduced annotation budget. To investigate this, we adapt existing Active Learning (AL) strategies to the task of fine-tuning PLM rankers and investigate their effectiveness, also considering annotation and computational costs. Our extensive analysis shows that AL strategies do not significantly outperform random selection of training subsets in terms of effectiveness. We further find that gains provided by AL strategies come at the expense of more assessments (thus higher annotation costs) and AL strategies underperform random selection when comparing effectiveness given a fixed annotation cost. 
Our results highlight that ``optimal'' subsets of training data that provide high effectiveness at low annotation cost do exist, but current mainstream AL strategies applied to PLM rankers are not capable of identifying them.
\vspace{-6pt} 
\end{abstract}


\keywords{\vspace{-4pt} \small PLM-based rankers, domain adaptation, active learning}

\maketitle

\vspace{-6pt}
\section{Introduction}
Search methods based on Pre-trained Language Models (PLM) have shown great effectiveness gains compared to common statistical models and early neural methods~\cite{lin2021pretrained,craswell2021trec,tonellotto2022lecture,formal2021splade,formal2021spladev2}. These language models are pre-trained for language representation learning on a background corpus; they are then further trained for a specific task -- a process commonly referred to as fine-tuning. Typically, PLM rankers are created through the fine-tuning of a PLM to the ranking task (and possibly, to a specific domain).
The fine-tuning of PLM rankers typically requires a great amount of labelled training data. This can often be a challenge when considering search tasks with no or little training data available. Data annotation typically requires a large manual effort and thus is expensive, especially in domain-specific tasks where annotators should be domain experts. In real-life settings, annotation and computational budget\footnote{With annotation budget we refer to the amount of money set aside for paying annotators to label pairs of queries and documents. With computational budget, we refer to the amount of money set aside for paying the computation costs arising from the training/fine-tuning of the PLM rankers. These costs may include the hardware and energy costs, or the purchase of cloud solutions.} is often limited, especially for start-ups or in domain-specific contexts.  

In this paper we focus on the problem of fine-tuning PLM rankers under limited training data and budget. There are alternative directions one may take to deploy a PLM ranker in a specific task for which no or limited training data is available. These include for example the zero-shot application of PLM rankers trained on another, resource-rich, retrieval task or domain~\cite{xin-etal-2022-zeroshot,thakur2021beir}, the learning with few-shot examples~\cite{dai2022promptagator}, and approaches based on pseudo-labelling~\cite{wang2021gpl}. However the effectiveness of these approaches depends on the relatedness of the fine-tuning task or the pre-training domain of the language model to the target retrieval task~\cite{wang2022neural}; thus their generalization capabilities remain unclear.
Therefore performing domain adaptation by fine-tuning the PLM ranker on the target task with annotated training data (the setting investigated in this paper) remains favourable for a (reliable) high effectiveness \cite{craswell2021benchmarkingmsmarco}.

It is unclear however how much annotated training data is required for training an effective PLM ranker. Furthermore, in presence of a budget constraint that restricts the amount of data that can be annotated for training, it is unclear whether it is possible to select training data to minimise annotation cost while maximising ranker effectiveness.

In this paper, (1) we investigate how the amount of labelled data used for fine-tuning a PLM ranker impacts its effectiveness, (2) we adapt active learning (AL) strategies to the task of training PLM rankers, (3) we propose a budget-aware evaluation schema including aspects of annotation and computation cost, (4) we conduct an extensive analysis of AL strategies for training PLM rankers investigating the trade-offs between effectiveness, annotation budget and computational budget. We do this in the context of three common PLM ranker architectures: cross-encoders (MonoBERT~\cite{nogueira2019bert}), single representation bi-encoders (DPR~\cite{karpukhin-etal-2020-dense}) and multi-representation bi-encoders (ColBERT~\cite{khattab2020colbert}), and  two scenarios:

\begin{itemize}[leftmargin=10pt]
	\item[\ding{202}] \textbf{Scratch}: the PLM is pre-trained on a background corpus, but has yet to be fine-tuned to the target  ranking task and dataset;
	\item[\ding{203}] \textbf{Re-Train}: domain adaptation of the PLM ranker is performed. The PLM is pre-trained on a background corpus and fine-tuned to a ranking task and a specific dataset, but further fine-tuning has yet to be performed to transfer the ranker to another dataset and, possibly, a ranking task with characteristics that differ from those of the first fine-tuning process.
\end{itemize}


To investigate the effect of the amount of labelled data on the effectiveness of PLM rankers, we select incremental amounts of data to fine-tune a ranker.
Our empirical results show that the size of the dataset available for fine-tuning the PLM ranker greatly influences the effectiveness of the ranker. While, somewhat unsurprisingly, we find that in general more training data leads to higher effectiveness, we also find large variability in effectiveness between different randomly selected training sets of the same size. Furthermore we find that, for some training sizes, the best random selection run outperforms the worst one, and significantly. This shows that there are subsets of the training data which lead to significant improvements within the same training data size.

This variability motivates us to investigate whether we can select those ``high-yield'' samples using Active Learning strategies. The intuition is that a good selection strategy would lead to a smaller amount of data to be annotated, and thus a lower annotation cost, while still producing a highly effective ranker.
Selection of training data has been extensively investigated in AL for machine learning. Here, common active selection strategies are based on uncertainty or diversity criteria \cite{CohnEtAl1996activelearning,lewis1994uncertaintysigir,shen2005aldiversity}. 
We thus adapt representative methods that implement these criteria to the context of fine-tuning PLM rankers. We evaluate the representative active selection strategies in terms of their effectiveness for fine-tuning PLM rankers on different training data sizes and compare the strategies to random selection of training data as baseline. 
For both scenarios the active selection strategies do not offer statistically significant improvements compared to random selection. For certain scenarios and PLM rankers we find varying beneficial selection strategies, however no selection strategy shows consistent and robust higher effectiveness than random selection. In addition, the adoption of active learning requires extra computation compared to random selection.

Since it is not our goal to minimize the training data size, but actually we aim to minimize the total cost of fine-tuning PLM rankers, we revisit the results in light of a budget-aware evaluation we introduce in this paper. This evaluation includes aspects of annotation cost as well as cost of computing resources. 
With this, we find that the annotations are the main cost factor. Since the selection methods require a different number of assessments to annotate a training set of a certain size, we compare the number of assessments to the effectiveness of the PLM rankers for random and active selection strategies.
This reveals that the (marginal, if any) effectiveness gains provided by AL strategies come at the expense of more assessments (thus higher annotation costs) and AL strategies under-perform random selection when comparing both effectiveness and associated cost.


%

We publish our code at: \textit{github.com/sophiaalthammer/al-rankers}.

%
\vspace{-6pt}
\section{Related work}
\label{sec:relatedwork}


\textbf{Effect of Data Size on PLM Rankers.}
Previous studies have observed that fine-tuning PLM rankers on subsets of the available training data decreases search effectiveness and, similarly, that increasing the size of the training data tends to improve search effectiveness. 
These types of observations and preliminary findings are reported for MS MARCO~\cite{karpukhin-etal-2020-dense,gupta2022survivorship,froebe2022traintestleakage,zhan2022interextrapolation,craswell2021benchmarkingmsmarco} and in the case of domain adaptation~\cite{hu2022p3ranker,gao-callan-2021-condenser,wang2021gpl,iurii2021systematicevaluatointransfer}.
However, these conditions have never been systematically evaluated, which we do in our study.  For example, \citet{nogueira-etal-2020-document} observe the variability in effectiveness when fine-tuning a MonoBERT ranker on subsets of different size (1k, 2.5k, 10k); however they do not systematically investigate this variability, nor they study the effect of larger subsets or different ranking architectures.
\citet{iurii2021systematicevaluatointransfer} investigate transfer learning for MonoBERT rankers first fine-tuned on MS MARCO and then transferred to question answering tasks in a zero-shot and full training setting, where the source and target domain hold large training sets. They also investigate the effect of training on subsets of the training data and find that the more training queries, the higher the effectiveness.
\citet{zhang-etal-2020-little} investigate domain transfer of BERT cross-encoders in a small data regime where they transfer MonoBERT from web search (trained on MS Marco) to small domain specific retrieval tasks. Interestingly they find that small in-domain training data sometimes decreases search effectiveness compared to the zero-shot application of MonoBERT.


\textbf{Active Learning for Information Retrieval.}
Active Learning aims to minimize the annotation cost associated with the acquisition of training labels while maximizing the effectiveness of the trained model. Uncertainty~\cite{lewis1994uncertaintysigir} and diversity-based~\cite{shen2005aldiversity} strategies form the bulk of AL methods that have been proposed and extensively validated across a variety of learning tasks and datasets. In this paper, we adapt methods belonging to these two strategies. 

Specifically, we explore the use of Active Learning for selecting data for the fine-tuning of PLM rankers. Active Learning has been used in Information Retrieval across a number of tasks and settings~\cite{CohnEtAl1996activelearning,LEWIS1994hetergenousuncertaintysampling,seung1992qbc,yang2009querysampling,shen2005activefeedback,zhu-etal-2008-activeuncertainty,long2015alforranking,donmez2009ecir,rodrigo2014alltr,donmez2008optimizing}, but never before in the context of PLM ranker fine-tuning. 

Of particular interest for this paper are the methods of \citet{cai2011rightteacher} and \citet{xu2007diversityal}, that we describe next, because we adapt them to our task of fine-tuning a PLM ranker. 
\citet{cai2011rightteacher} transfer a learning-to-rank (LTR) model trained on one target domain to another source domain. For the domain adaptation training they propose to use the Query-by-Committee algorithm for active selection of queries in the target and in the source domain as well as mixing the training sets of the target and source domain. For the domain adaptation of a LTR model, QBC reaches a higher retrieval effectiveness with less training data than the random selection strategy. 
\citet{xu2007diversityal} investigate different diversity-based active learning strategies for updating query relevance scoring and propose a combination of diversity and density based selection.

A variation of the AL setting that has shown success in certain domain-specific tasks is that of continuous active learning~\cite{Grossman2011TechnologyAssistedRI,yang2022goldilocks,sadri2022albert}, where documents are iteratively retrieved by actively learning for one specific query, typically aiming for total recall~\cite{trec2016totalrecall}. For the task of technology assisted review (TAR), \citet{yang2021minimizingcost} propose a TAR cost framework, however this framework focuses on cost modeling for reviewing one specific query.

Despite previous successes in the use of AL strategies in the context of search and ranking, AL strategies have not been studied for PLM rankers. The AL strategies proposed in previous work are not directly applicable to PLM rankers -- however in Section \ref{sec:activeselectionstrategies} we propose adaptations of these methods to our task of interest.

\vspace{-6pt}
\section{Considered PLM rankers} \label{sec:PLM-Considered rankers}


In the consider cross-encoder model, MonoBERT, query and passage text are concatenated, encoded with BERT and the CLS representation is scored with a linear layer $W$ on top of the encoding:
\begin{equation}
	\begin{aligned}
		s = W \ \text{BERT}(\text{CLS};q;\text{SEP};p;\text{SEP})_{CLS} 
	\end{aligned}
\end{equation}
\noindent where SEP is the separator token and $s$ is the final score of passage $p$ for query $q$.
Empirical findings show MonoBERT reaches a high re-ranking effectiveness~\cite{nogueira2019bert}, however each passage needs to be encoded at query time and therefore this architecture is computationally resource-heavy and is characterized by high query latency~\cite{scells2022reduce}. For the same reason, this ranker is commonly used only in top-$k$ re-ranking settings, and not for retrieval (i.e., scoring the whole collection for each query).


DPR~\cite{karpukhin-etal-2020-dense} encodes the query and passages independently. The relevance of a passage $p$ to a query $q$ is estimated using the dot-product between the CLS token representation $q$  and that of $p$:
\begin{equation}
\begin{aligned}
s = \text{BERT}(\text{CLS};q;\text{SEP})_{CLS} \cdot \text{BERT}(\text{CLS};p;\text{SEP})_{CLS}
\end{aligned}
\end{equation}
The independence of query and passage encoding and dot-product relevance scoring make it possible to pre-compute and store the passage representations in the index and enable efficient retrieval at query time with approximate nearest neighbor search~\cite{malkov2016hnsw,johnson2019billionfaiss}.

The ColBERT~\cite{khattab2020colbert} method delays the interaction between the query and passage to after the encoding by computing the relevance score as the sum of the maximum similarity scores between all token representations of the query and passage:
\begin{equation}
\begin{aligned}
s = \sum_{j} \max_{i} \big [\text{BERT}(\text{CLS};q;\text{SEP})_j \cdot \text{BERT}(\text{CLS};p;\text{SEP})_i \big ]
\end{aligned}
\end{equation}
As for single representation bi-encoder methods, also in ColBERT the passage representation can be pre-computed offline and thus the query processing is sped up. Empirical results show that ColBERT achieves a competitive effectiveness compared to MonoBERT~\cite{khattab2020colbert}.

\vspace{-6pt}
\section{Training Scenarios \& Annotation Modeling}
\label{sec:trainscenarios}

We consider two scenarios for training the PLM rankers: \ding{202} \textbf{Scratch} training from scratch, starting with a PLM and \ding{203} \textbf{Re-Train} domain fine-tuning after rank/retrieval fine-tuning of the PLM has occurred. These are common scenarios that are encountered in the practical application of PLM rankers to search problems. 

In \ding{202} \textbf{Scratch} our objective is to train a PLM ranker ``from scratch'', i.e., without having already performed any fine-tuning on a retrieval task.
There are many reasons this scenario could occur in the practical deployment of PLM rankers. For example, no suitable labelled data corresponding to the ranking task may be available, or the data that may be available is protected by a license that prevents its use within a product (e.g., the MS Marco dataset).
We model the first scenario by starting from a pre-trained BERT model \cite{devlin-etal-2019-bert,sanh2019distilbert} and training the ranker on the MS Marco dataset, a large scale web search collection commonly used to train these rankers. Note, in our experiments we assume that no labels are available for the dataset, and labels are iteratively collected (in a simulated setting) within the AL cycle.

In \ding{203} \textbf{Re-Train} our goal is to adapt a PLM ranker to a specific retrieval task (potentially in a specific domain). Here we assume that the PLM ranker has already undergone fine-tuning on a high-resource retrieval task (e.g., using the common MS Marco dataset), and the goal is to further fine-tune the ranker with additional data, on a different retrieval task or data domain. This is a common setting in domain-specific IR settings. The assumption is that the initial fine-tuning on the non-target retrieval task or domain data still highly contributes to the effectiveness of the ranker, especially when the target data available for fine-tuning is limited.
We model the second scenario by starting from a ranker fine-tuned on MS Marco and fine-tune the ranker for a domain-specific retrieval task. 
In our experiments, we choose to validate the models using the retrieval task and datasets associated with health-oriented web search in the medical domain.
We choose this task due to the availability of the TripClick dataset \cite{rekabsaz2021tripclick}, a large-scale training and test set for this task. This dataset has similar characteristics to MS Marco (e.g. query length, sparse judgments). 
In contrast to other domain adaptation approaches~\cite{cai2011rightteacher}, we do not mix the training sets of the source and the target domain, in order to (i) be able to separate the effects of mixing the training sets from the active domain adaptation strategies, and (ii) study PLM ranker development and deployment strategies that are in line with the green IR principles of reuse and recycle~\cite{scells2022reduce}.

In order to model the real-life process of incremental annotation and training we incrementally increase our fine-tuning set $D$. The details of this incremental process are depicted in Algorithm \ref{fig:algorithm}. We start with an empty set $D = \{\}$ and in each iteration a subset $S$ of the whole training set $T$ ($S \subset T$) is selected to be added to $D$. We model the annotation process by attaining the labels from the training set qrels and adding the samples to the fine-tuning set ($D = D \cup S$). Then we train the PLM ranker on the updated set $D$ and, based on random or active selection strategies, we select the next subset to annotate and add it to the training set.

\vspace{-6pt}
\section{Active Selection Strategies}
\label{sec:activeselectionstrategies}

We consider three active selection strategies to identify training data for labelling: uncertainty-based selection \cite{lewis1994uncertaintysigir,zhu-etal-2008-activeuncertainty,yu2005svmselectivesampling}, query-by-committee (QBC) \cite{cai2011rightteacher}, and diversity-based selection~\cite{xu2007diversityal}. We consider random selection as a baseline selection strategy. Next, we describe the active selection strategies and how we adapt them for fine-tuning PLM rankers.


\vspace{-10pt}
\subsection{Uncertainty-based selection}
The uncertainty-based selection strategy selects samples by measuring the model's (ranker) uncertainty in the scores it produced and then selecting the samples with the least confidence~\cite{lewis1994uncertaintysigir,CohnEtAl1996activelearning}. Uncertainty-based strategies are commonly applied to classification problems, and often the score provided by the classifier is used as direct indication of uncertainty: scores are in the range $[0, 1]$, the decision boundary is set to $0.5$ and the confidence in the classification is measured in function of the distance to the decision boundary (the closer, the least confident)~\cite{ein-dor-etal-2020-active,lewis1994uncertaintysigir}. 

This approach is not directly transferable to PLM rankers since their relevance scores are not necessarily bounded and therefore there is no clear decision boundary measuring the uncertainty in the ranking.
We note that uncertainty estimation in Information Retrieval is a fundamental but largely unexplored problem~\cite{turtle1997uncertainty,crestani1998information,collins2007estimation}, especially for rankers based on PLMS~\cite{lesota2021modern,cohen2021not}. 

In this work, we model the uncertainty distribution by means of the score distribution of the top $K$ ranked passages for all queries in the training set $T \textbackslash D$ and select the query-passage pairs with the relevance score closest to the mean of the score distribution for the ranker, hence with the highest uncertainty. We leave the investigation of other upcoming approaches for future work (see Section~\ref{conclusions} for further insights).

In order to model the annotation and training process for the selected query-passage pairs, the selected passage is assigned its label from the training set. In case the selected passage is \textit{relevant} we sample an irrelevant passage randomly from the BM25 top 1,000 to construct a training triplet (query, positive passage, negative passage). In case the selected passage is \textit{irrelevant}, we take the selected passage as negative for the triplet and take the first relevant passage in the BM25 top100 list re-ranked by the PLM ranker as positive passage. For each selected query-passage pair we add one triplet to the training set.

Note that the uncertainty-based selection is operating at query-passage level, while the other AL methods we consider next are operating at a query-only level.

\vspace{-10pt}
\subsection{Query-by-committee selection}

The query-by-committee (QBC) method~\cite{seung1992qbc,freund1997qbc} is a specific uncertainty-based selection strategy. 
In QBC, multiple committee members (models) are used to classify or rank samples; then the disagreement between the committee members on classified/ranked samples is measured and the samples with the highest disagreement are selected for annotation. A previous adaptation of QBC to information retrieval is due to \citet{cai2011rightteacher} who apply QBC to Learning-to-Rank for domain adaptation. 
For this, they train different members of the committee by training on subsets of the currently annotated training set at hand. The disagreement between the committee members is then measured by the vote entropy of the different rankings of the members for the queries which are not yet annotated. In the vote entropy the committee members $M$ vote on the partial order of two passages $N(p_1 \prec p_2)$ in the ranked list $R$, counting how many of the members rank $p_1$ higher than $p_2$. The vote entropy of a query $q$ is then defined as:

\begin{algorithm}[t]
\caption{Incremental annotation and training process}\label{alg:cap}
\label{fig:algorithm}
\begin{algorithmic}
\Require $T$ whole training set, $I$ number of iterations, $s$ number of added samples per iteration, $M$ is a PLM ranker/retriever 
\Ensure $D$ annotated training set, $M$ PLM ranker trained on $D$
\State $D \gets \{\}$
\For{$i$ in $I$}
\State Select subset $S \subset T$ of size $|S|=s$ with selection strategy
\State Annotate $S$, $T \gets T \setminus S$ 
\State $D \gets D \cup S$
\State Train $M$ with $D$
\EndFor
\end{algorithmic}
\end{algorithm}

\vspace{-12pt}
\begin{equation}
	\vspace{-4pt}
VE(q) = \frac{-1}{|M|} \sum_{p_i, p_j \in R} N(p_i \prec p_j) log \left( \frac{N(p_i \prec p_j)}{|M|} \right)
\end{equation}

The queries with the highest vote entropy are selected for annotation.
In order to model the annotation process and to select training triples for training PLM rankers, for every query that selected to add to the training set we take the first relevant passage in the BM25 top100 list re-ranked by the PLM ranker as positive passage and sample a random negative passage from the BM25 top 1,000 passages. We choose to use the re-ranked list of the first member for selection. For every query we add one training triplet to the training set.

\subsection{Diversity-based selection}
\label{sec:budgetaware}

Diversity-based selection strategies select training samples based on the diversity of the samples -- typically comparing already selected samples to those yet to select. Within Information Retrieval, diversity-based selection strategies have been used for Learning-to-Rank models \cite{xu2007diversityal,shen2005aldiversity,yang2009querysampling}.
In those settings, diversity is measured by clustering queries using an external unsupervised clustering model and taking one representative query from each cluster.

In our adaptation of diversity-based selection strategies to PLM rankers, to compute diversity we consider the query representation made by the PLM ranker. 
This has the advantage that we leverage the model's representations to compute diversity, instead of relying on an external model. 
Furthermore, this representation changes in each iteration as the PLM ranker is trained  incrementally through training sets of increasing size: therefore, the query representation also accounts for changes within the ranker itself.
For DPR and ColBERT, we use the CLS token representation of the encoded query as query representation. For MonoBERT we encode the query without a passage and also take the CLS representation for measuring the diversity.
We cluster the query representations of queries in $T \textbackslash D$ with the number of clusters equaling the number of training samples to be added in that iteration. From each cluster, we randomly sample one query to be annotated.

To add a training triplet to the training set for each selected query, we rely on the same annotation process used in QBC.

\section{Budget-aware evaluation}
\label{sec:costmethodology}

Next we introduce a framework to evaluate active learning to PLM ranker fine-tuning within budget constraints. For this, we model the costs related to both annotation effort and computation.

\subsubsection*{Annotation Costs}

For measuring the annotation costs, we count the number of assessments needed to annotate a training triplet for a single query. The number of assessments corresponds to the rank of the first relevant passage found in the ranking of the PLM ranker, which is trained in the previous iteration of the AL process. In our setting, the annotation for a query stops when one relevant passage for a selected query is found -- thus we need to assess all passages in the ranking for that query up until the relevant passage is found and also annotated.This implies that the number of assessments differs from the training data size (the number of queries), e.g., one query sample in the training data can account for 10 assessments required when the first relevant passage is found at rank 10.

The total annotation cost then is the sum of the number of assessments for all the queries added to the training set. Formally, let $A(i)$ be the number of assessments needed to create the training data of iteration $i$, $A_h$ be the number of assessments an annotator finishes in one hour and $A_C$ be the cost for an annotator per hour\footnote{Note that certain search tasks or domain may require multiple annotators to examine the same sample: in this case $A_C$ would be the sum of the hourly rates associated to all the annotators.}. Then, the total annotation cost at iteration $i$ is computed as:

\vspace{-10pt}
\begin{equation}
	\vspace{-4pt}
	C_A(i) = \frac{A(i)}{A_h} \cdot A_C \label{annotation-costs}
\end{equation}

\subsubsection*{Computational Costs}
Next we model the computational costs involved in executing the active selection strategies. These strategies usually will require both CPU and GPU based computation, which typically incur different costs and thus we account for separately. 
Let $H_{GPU}(i)$ be the accumulated number of GPU hours needed for training a PLM ranker for iteration $i$ and $G_{h}$ be the cost of running an GPU for one hour. Then, the total computational cost at iteration $i$ is computed as:

\vspace{-12pt}
\begin{equation}
	\vspace{-2pt}
	C_C(i) = H_{GPU}(i) \cdot G_h +H_{CPU} \cdot C_h \cdot (i-1) \label{compute-costs}
\end{equation}

\noindent with $H_{CPU}$ the number of CPU hours needed for computing the selection strategy and $C_h$ the cost of one hour CPU.

\vspace{-4pt}
\subsubsection*{Total Cost}


Finally, the total cost at iteration $i$ can then be computed using Equations~\ref{annotation-costs} and~\ref{compute-costs}:

\vspace{-10pt}
\begin{eqnarray}
	\vspace{-10pt}
    &C(i) &= C_A(i) + C_C(i) \\
    &&= \frac{A(i)}{A_h} \cdot A_C +  H_{GPU}(i) \cdot G_h +H_{CPU} \cdot C_h \cdot (i-1) \nonumber
\end{eqnarray}

\section{Experimental Setup}
\label{sec:experiments}
Next we describe the experimental setup we have devised to study the AL strategy for PLM rankers fine-tuning we have illustrated above. We develop our investigation along the following three lines of inquiry:


\newcommand{\RQone}{\begin{itemize}
    \item[\textbf{RQ1}] What is the effect of the size of the labelled training data on the effectiveness of PLM rankers? 
\end{itemize}}
\RQone

\newcommand{\RQtwo}{\begin{itemize}
    \item[\textbf{RQ2}] How do different active selection strategies influence the effectiveness of PLM rankers?
\end{itemize}}
\RQtwo

\newcommand{\RQthree}{\begin{itemize}
    \item[\textbf{RQ3}] What is the effect of using an active selection strategy to fine-tune a PLM ranker under a constrained budget?
\end{itemize}}
\RQthree

\subsection{Passage Collection \& Query Sets}
For \ding{202} \textbf{Scratch}, we use the MS Marco passage collection \cite{msmarco16}. MS Marco is based on sampled Bing queries and contains $8.8$ million passages; its training set contains $530k$ training triplets. 
We use the training portion for fine-tuning and evaluate on the TREC DL 2019~\cite{trec2019overview} and 2020~\cite{trec2020overview} with nDCG@10.

For \ding{203} \textbf{Re-Train}, we use the TripClick dataset. This dataset contains real user queries and click-based annotations. 
It consists of $1.5$ million passages and $680k$ training queries. Test queries are divided with respect to their frequency into three sets of $1,750$ queries respectively; the three sets are Head, Torso, and Tail. For the Head queries a DCTR~\cite{chuklin2015click} click model was used to create relevance signals from the click labels. We evaluate on the Head DCTR and the Torso Raw test set as in related work \cite{rekabsaz2021tripclick, hofstaetter2022tripclick,althammer2022tripjudge}.

\vspace{-8pt}
\subsection{PLM ranker details}
\vspace{-2pt}
We train MonoBERT, ColBERT and DPR using training triplets with a RankNet loss \cite{burges2010ranknet}. The triplets consist of the query, a relevant and an irrelevant passage; negative passages are taken from the top 1000 BM25 negatives. We train DPR and ColBERT with a batch size of $100$, while we use a batch size of $32$ for MonoBERT due to its high computational requirements.
We train all models for $200$ epochs with a learning rate of $7 \times 10^{-6}$ and we use early stopping.
For training, we impose a maximum input length of $30$ tokens for the query and $200$ tokens for the passage; this setting truncates only a few outliers samples in the dataset but provides computational advantages for batching.

In \ding{202} \textbf{Scratch} we perform fine-tuning from scratch; as underlying PLMs we use DistilBERT~\cite{sanh2019distilbert} for DPR and ColBERT and the \texttt{bert-base-uncased} model~\cite{devlin-etal-2019-bert} for MonoBERT both provided by Huggingface. We choose these models as starting point so that they match the fine-tuned models for \ding{203} \textbf{Re-Train}. In \ding{203} \textbf{Re-Train} we start with PLM rankers fine-tuned on MS Marco. For DPR we start from TASB~\cite{Hofstaetter2021_tasb_dense_retrieval}, trained with knowledge distillation and topic-aware sampling; for ColBERT from a ColBERT DistilBERT model trained with knowledge distillation; for MonoBERT from a \texttt{bert-base-uncased} model solely trained on MS Marco.

For MonoBERT and ColBERT, we report results in a re-ranking context, i.e. using these PLM rankers to re-rank the top 1,000 results retrieved by BM25. For DPR, we instead consider a retrieval setting, where all the collection is scored and then only the top 1,000 are used for evaluation. However, the findings we observe for DPR in the retrieval setting are similar to those we obtained for the same PLM in a re-ranking setting (not reported here). We decided to report retrieval results for DPR, rather than re-ranking as for the other two PLM, because DPR is more commonly used for retrieval (while the other two for re-ranking).

\vspace{-8pt}
\subsection{Active learning details}
As foundational experiment we train the PLM rankers on different subsets of the training data of differing sizes; as size, we explore the values $[1k, 5k, 10k, 20k, 50k, 100k, 200k]$ for MS Marco and $[1k, 5k, 10k, 20k, 50k]$ for TripClick. We repeat these experiments 4 times with different random seeds for sampling the subsets, so that each time we train on different subsets with the same size and we can measure variance.

In our experiments we use random selection as a baseline and increase the training set incrementally. We run the random baseline 4 times with different random seeds.

For the active learning process, we increase the training subsets incrementally as denoted in Algorithm 1. In each iteration we train the PLM ranker from scratch to exclude a potential bias from incrementally training a ranker. In the first iteration we randomly select the first subset with the same random selection across the different active learning strategies. For uncertainty and diversity selection one could select the first batch with the selection strategy, however for QBC this is not possible because different committee members for selection are not available in the first iteration. Therefore we do random selection in the first iteration to be able to fairly compare across the three strategies.

For fast and resource efficient active selection, we train the PLM rankers for $15$ epochs and use the trained ranker for active selection. For the sake of evaluation and in order to compare the effectiveness at different iterations, we resume the training after $200$ epochs.

For the uncertainty-selection and QBC strategies, we score the BM25 top 100 passages of each training queries and use these passages for actively selecting the queries for annotation.
For the QBC selection strategy we use the same hyper-parameters as \citet{cai2011rightteacher}; we use $2$ members in the committee and train each member on $80\%$ of the subset available at each iteration for training. We choose the size of the training subsets so that each $80\%$ portion aligns with the other training set sizes.

For \ding{202} \textbf{Scratch} we add in each iteration $5,000$ training samples to the training size. For \ding{203} \textbf{Re-Train} we have $5,000$ samples for the first 2 iterations until the training size is 10k, from then on we use $10,000$ samples for the remaining iterations in order to decrease computational cost.

\vspace{-8pt}
\subsection{Costs for Budget-Aware Evaluation}
\vspace{-4pt}
For computing the annotation cost, for each triplet added to the training we store the rank of the first relevant document in the ranked list generated by the PLM ranker trained in the previous iteration. Since we do not have a trained PLM ranker in the first iteration, we start with the initial ranking provided by BM25. For the random baseline, we also use the initial BM25 ranking for computing the annotation effort. 

We conduct our experiments on servers equipped with NVIDIA A40 GPUs and measure the GPU and CPU hours spent in the training of the PLM ranker and the execution of the selection strategies.

For the computational cost, we refer to common cloud computing costs\footnote{From \url{https://aws.amazon.com/ec2/pricing/on-demand/}. Costs valid as of 02 January 2023. GPU costs refer to a \texttt{p3.2xlarge} instance and CPU costs to an \texttt{a1.4xlarge} instance.} and set $G_{h}=3.060\$$ and $C_{h}=0.408\$$.
For the number of annotations per hour $A_h$ we rely on estimates from \citet{althammer2022tripjudge} who conducted an annotation campaign on TripClick test set. Here annotators needed $47.7$ seconds to annotate a query-passage pair on average, which corresponds to $75$ assessments per hour. For the annotation cost per hour, $A_C$, we assume $50US\$$ as hourly rate of a domain expert annotator. We also have developed a small web tool\footnote{We make this publicly available on acceptance of the paper.} to let the reader customise computation and annotation costs and the number of assessments per hour; then the reader can observe how the results presented in terms of budget-aware evaluation change according to different cost settings.

\vspace{-6pt}
\section{Results}
\label{sec:results}


%

\begin{figure}[t!]
	\centering
	\captionsetup[subfloat]{farskip=0.5pt,captionskip=0.5pt}
	\subfloat[\ding{202} \textbf{Scratch}. \label{fig:boxplot-scratch}]{\includegraphics[width=0.5\linewidth]{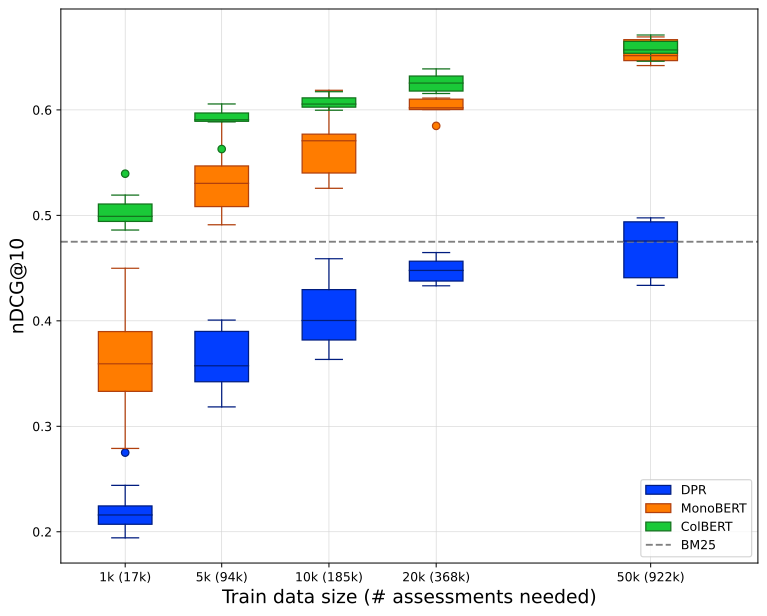}}
	\subfloat[\ding{203} \textbf{Re-Train}. \label{fig:boxplot-retrain}]{\includegraphics[width=0.5\linewidth]{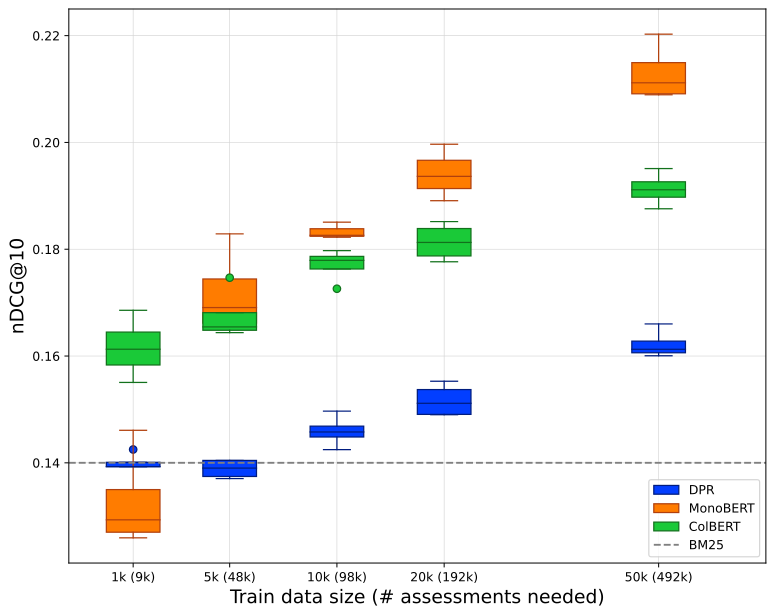}}
	\caption{Boxplot of nDCG@10 effectiveness on TREC DL 2020 (\ding{202} \textbf{Scratch}, Figure~\ref{fig:boxplot-scratch}) and on TripClick Head DCTR test (\ding{203} \textbf{Re-Train}, Figure~\ref{fig:boxplot-retrain}), visualizing the variability of training on different training sample sizes. PLM rankers are trained on subsets of respective sets (MS Marco/TripClick) with different sizes. To measure variability, for each train data size we repeat random sampling 4 times.}
\end{figure}

\vspace{-2pt}
\subsection{RQ1: Effect of Size of Training Data}

We visualize the effect of training data size on the effectiveness of PLM rankers for \ding{202} \textbf{Scratch} (Figure \ref{fig:boxplot-scratch}) and for \ding{203} \textbf{Re-Train} (Figure \ref{fig:boxplot-retrain}). The boxplots visualize the range of effectiveness when the PLM ranker is trained on different subsets of the same size.

In both cases, it is observed that as the size of the training data increases, nDCG@10 improves for all three PLM rankers. When considering effectiveness across PLM rankers, it is noteworthy to observe ColBERT and MonoBERT. Recall from the literature that MonoBERT outperforms ColBERT on MS Marco when both are trained on the whole MS Marco training data \cite{craswell2021trec}, and the same holds for TripClick~\cite{hofstaetter2022tripclick}. However, in our experiments, we observe that ColBERT outperforms MonoBERT for smaller training data sizes. MonoBERT eventually becomes better than ColBERT but only once more than 10,000 training samples are used for the \ding{203} \textbf{Re-Train} scenario. For the \ding{202} \textbf{Scratch} scenario the two rankers becomes largely indistinguishable when the training data is 50,000 samples, and eventually MonoBERT takes the lead thereafter (not shown in the figure).

\begin{table*}[t]
	\small
	\centering
	\caption{nDCG@10 effectiveness across different amounts of training data for \ding{202} \textbf{Scratch} on TREC DL 2019 \& 2020 and for \ding{203} \textbf{Re-Train} on TripClick Head DCTR \& Torso Raw. Bold numbers denote highest effectiveness for each PLM ranker and training size. Statistical significant differences to random selection baseline (Random) 
		is denoted with * (paired t-test; p < 0.05, Bonferroni correction with n=3). For \ding{202}: No consistently best performing method and no statistical significant difference to Random. For \ding{203}: for DPR, Random consistently is best; all statistical significance differences to Random are significantly lower. `-' indicates no result at that training size.}
	
	\label{tab:eff}
	
	\begin{adjustbox}{width=0.95\textwidth, center=\textwidth}
	\setlength\tabcolsep{2.0pt}
	\begin{tabular}{cl!{\color{black}\vrule}l!{\color{lightgray}\vrule}l!{\color{lightgray}\vrule}l!{\color{lightgray}\vrule}l!{\color{lightgray}\vrule}l!{\color{black}\vrule}l!{\color{lightgray}\vrule}l!{\color{lightgray}\vrule}l!{\color{lightgray}\vrule}l!{\color{lightgray}\vrule}l!{\color{black}\vrule}l!{\color{lightgray}\vrule}l!{\color{lightgray}\vrule}l!{\color{lightgray}\vrule}l!{\color{lightgray}\vrule}l!{\color{black}\vrule}l!{\color{lightgray}\vrule}l!{\color{lightgray}\vrule}l!{\color{lightgray}\vrule}l!{\color{lightgray}\vrule}l}
		\toprule
		
		
		\multicolumn{2}{c!{\color{black}\vrule}}{\textbf{nDCG@10}}&\multicolumn{10}{c!{\color{black}\vrule}}{\textbf{\ding{202} Scratch: MS Marco}}&
		\multicolumn{10}{c}{\textbf{\ding{203} Re-Train: TripClick}}\\
		
		\multicolumn{2}{c!{\color{black}\vrule}}{}&\multicolumn{5}{c!{\color{black}\vrule}}{\textbf{TREC DL 2019}}&
		\multicolumn{5}{c!{\color{black}\vrule}}{\textbf{TREC DL 2020}}&\multicolumn{5}{c!{\color{black}\vrule}}{\textbf{Head test DCTR}}&
		\multicolumn{5}{c}{\textbf{Torso test raw}}\\
		
		\multicolumn{2}{c!{\color{black}\vrule}}{\textbf{Train data size}} &
		\multicolumn{1}{c!{\color{lightgray}\vrule}}{\textbf{0}} &
		\multicolumn{1}{c!{\color{lightgray}\vrule}}{\textbf{5k}} &
		\multicolumn{1}{c!{\color{lightgray}\vrule}}{\textbf{10k}}&
		\multicolumn{1}{c!{\color{lightgray}\vrule}}{\textbf{20k}}&
		\multicolumn{1}{c!{\color{black}\vrule}}{\textbf{50k}}&
		\multicolumn{1}{c!{\color{lightgray}\vrule}}{\textbf{0}} &
		\multicolumn{1}{c!{\color{lightgray}\vrule}}{\textbf{5k}} &
		\multicolumn{1}{c!{\color{lightgray}\vrule}}{\textbf{10k}}&
		\multicolumn{1}{c!{\color{lightgray}\vrule}}{\textbf{20k}}&
		\multicolumn{1}{c!{\color{black}\vrule}}{\textbf{50k}} &
		\multicolumn{1}{c!{\color{lightgray}\vrule}}{\textbf{0}} &
		\multicolumn{1}{c!{\color{lightgray}\vrule}}{\textbf{5k}} &
		\multicolumn{1}{c!{\color{lightgray}\vrule}}{\textbf{10k}}&
		\multicolumn{1}{c!{\color{lightgray}\vrule}}{\textbf{20k}}&
		\multicolumn{1}{c!{\color{black}\vrule}}{\textbf{50k}}&
		\multicolumn{1}{c!{\color{lightgray}\vrule}}{\textbf{0}} &
		\multicolumn{1}{c!{\color{lightgray}\vrule}}{\textbf{5k}} &
		\multicolumn{1}{c!{\color{lightgray}\vrule}}{\textbf{10k}}&
		\multicolumn{1}{c!{\color{lightgray}\vrule}}{\textbf{20k}}&
		\multicolumn{1}{c}{\textbf{50k}}\\
		\midrule
		\arrayrulecolor{black}
		\textcolor{gray}{0}& BM25&.501&&&&&.475&&&&&.140&&&&&.206&&&\\
		
		\midrule
		\arrayrulecolor{lightgray}
		\multicolumn{10}{l}{\textbf{MonoBERT (re-rank BM25 top 1,000)}} &&&\\
		
		\textcolor{gray}{1} & Random &.051&.5935& .6272& .6430& .6705&   .041&.5590& .5871& .6148& .6552
		&.036&.1715&.1833&.1941&.2129 &.036&.2279&.2352&.2426&\textbf{.2710}\\
		\textcolor{gray}{2} & QBC &&\textbf{.6193}& .6157& .6246& \textbf{.6728}& & .5507& .5844& \textbf{.6443}& .6630
		&&\textbf{.1731}&.1835&\textbf{.2065}&.2059&&.2046&.2328&.2423 &.2679\\
		\textcolor{gray}{4} & Uncertainty && .6118& .6232& \textbf{.6588}&  .6509&   & \textbf{.5875}& .5873& .6336& .6595
		&&-&\textbf{.1920}&.1981&\textbf{.2190}&&-&\textbf{.2356}&.2362&.2705\\
		\textcolor{gray}{5} & Diversity &&.5925& \textbf{.6341}&  .6448& .6640 &  &.5407& \textbf{.6237}& .6338& \textbf{.6670}
		&&-&.1837&.1933&.2123&&-&.2294&\textbf{.2450}&.2650\\
		\midrule
		\arrayrulecolor{lightgray}
		\multicolumn{10}{l}{\textbf{ColBERT (re-rank BM25 top 1,000)}} &&&\\
		
		\textcolor{gray}{6} & Random &.352&.6176&\textbf{.6385}&.6352&.6614 &.246&.5944&.6091&.6291&.6577
		&.155&\textbf{.1675}&.1770&.1813&.1912 &.227&\textbf{.2300}&\textbf{.2351}&\textbf{.2397}&.2475\\
		\textcolor{gray}{7} & QBC  &&.6192&.6297&\textbf{.6541}&\textbf{.6680} &&.5813&.6159&\textbf{.6511}&\textbf{.6758}
		&&$.1302^{*}$&\textbf{.1791}&\textbf{.1860}&\textbf{.1962}& &$.1558^{*}$&.2273&.2292&.2360\\
		\textcolor{gray}{9} & Uncertainty &&.6257&.6034&.6089&.6370& &\textbf{.5987}&.6076&.6001&.6246
		&&-&.1645&.1536&$.1753^{*}$ &&-&.2190&.1909&.2274\\
		\textcolor{gray}{10} & Diversity &&\textbf{.6271}&.6239&.6402&.6644& &.5912&.6038&.6211&.6363
		&&-&$.1645$&.1811&.1957&&-&.2187&.2362&\textbf{.2481}\\
		
		\midrule
		\arrayrulecolor{lightgray}
		\multicolumn{10}{l}{\textbf{DPR (full retrieval)}} &&&\\
		
		\textcolor{gray}{11} & Random &0.0&.3674&\textbf{.4390}&.4457&.5006 &0.0&.3225&.3789&.4190&.4757
		&.139&\textbf{.1389}&\textbf{.1459}&\textbf{.1516}&\textbf{.1621} &\textbf{.200}&\textbf{.1837}&\textbf{.1745}&\textbf{.1924}&\textbf{.2023}\\
		\textcolor{gray}{12} & QBC &&.3465&.4343&.4628&\textbf{.5079}& &.3023&.3849&.4090&.4534
		&&$.0849^{*}$&.1043&.1368&.1603 &&.0895&.1122&.1312&.1440\\
		\textcolor{gray}{14} & Uncertainty &&\textbf{.3961}&.4067&.4255&$.3757^{*}$& &\textbf{.3660}&.3733&\textbf{.4476}&.4254
		&&.1060&.1165&.1283&.1336& &.0907&.0946&.1030&.1031\\
		\textcolor{gray}{15} & Diversity &&.3713&.4086&.4593&.4750& &.3437&.4030&.4198&\textbf{.4998}
		&&.1059&.1150&.1163&.1458&&.0907&.1041&.1217&.1473\\
		\arrayrulecolor{black}
		\bottomrule
	\end{tabular}
	\end{adjustbox}
	\vspace{-12pt}
\end{table*}


In \ding{202} \textbf{Scratch}, the improvement in effectiveness with increasing training size is particularly remarkable for small training subsets.
For example, the improvement by adding 4k training samples from 1k to 5k samples is between 18\% and 63\% of the median nDCG@10.
Noting the wide scale of the y-axis from 0.2 to 0.7 nDCG@10, we observe a large variability when training PLM rankers on limited data. This is particularly the case for MonoBERT, where we find a large difference from maximum and minimum nDCG@10 from 19 (0.27--0.46 nDCG@10) for 1k samples to to 7 (0.53--0.60) for 10k samples. The worst and the best MonoBERT run obtained are statistically different for train size 1k and 5k.
For DPR, the inter-quartile range is up to a difference of 5 nDCG@10 (0.38-0.43 for 10k), thus 50\% of the effectiveness points are within a range of 5 nDCG@10.
A substantial variability in the effectiveness of DPR is observed when trained on 50k samples. The best and the worst runs for DPR are statistically different for 5k and 10k samples.
It is noteworthy that the boxplots for 5k samples overlap in part with those for 10k, and similarly the 10k with those for 20k. This means that specific subset of training data of size 5k (10k) allow to reach the same effectiveness obtained when training the ranker on double the amount of data, i.e. 10k (20k).

For \ding{203} \textbf{Re-Train} (Figure \ref{fig:boxplot-retrain}) we also notice variability in search effectiveness; yet, we observe a relatively smaller variability compared to \ding{202} \textbf{Scratch}. The differences between the worst and best runs for each training data size are not statistically significant in this scenario. We suspect that this smaller variability in effectiveness is due to starting from an already fine-tuned PLM ranker instead of training from scratch.
Although our empirical results suggest a smaller variability, we still see overlaps of the boxplots, especially between 10k and 20k sample: that is, the same or even better effectiveness could have been reached with half the training data.

These results suggest that it is possible to select subsets of training data that would ``speed-up'' the learning: in other words, some subsets of training data can achieve the same or even higher effectiveness as using double the amount of data. This thus serves as a motivation for this paper: is it possible to identify ``\textit{high-yield}'' training subsets so as to spare annotation costs but yet obtain high effectiveness? To this aim, we investigate the effectiveness of active learning strategies, which we discuss next.


\vspace{-14pt}
\subsection{RQ2: Effectiveness of Active Selection}

In Table \ref{tab:eff} we report the effectiveness of the active learning strategies from Section \ref{sec:activeselectionstrategies}, along with the random selection baseline, when used for training MonoBERT, ColBERT and DPR across different amounts of training data in scenario \ding{202} \textbf{Scratch} and \ding{203} \textbf{Re-Train}.

For the random selection baseline we report the mean effectiveness when randomly sampling and training on different subsets of the same size multiple times -- we perform four random selections for each training size. Note that because the AL selection strategies are deterministic,  there is only one result for each strategy at a certain training size, not multiple runs as for the Random baseline.

Various AL strategies outperform the Random baseline at most training data sizes in \ding{202} \textbf{Scratch}. However these effectiveness gains are not consistent throughout all training sizes: there is no single AL strategy that always perform better than the others and, importantly, that always outperforms the random selection baseline.
For example, on TREC DL 2020 the uncertainty-based selection for DPR reaches the highest effectiveness when training with 20k samples, but effectiveness drops sensibly when training with 50k samples.
Furthermore, effectiveness gains across all methods are not statistically significant, nor are the improvements substantial. When evaluating the PLM rankers on MS Marco Dev, we find similar results, there are varying, non-statistically significant improvements of AL strategies to the Random baseline; due to space constraints we do not report these measures here.

\begin{figure*}[t]
\centering
\captionsetup[subfloat]{farskip=0.5pt,captionskip=0.5pt}

\resizebox{1\linewidth}{!}{

\begin{minipage}{1\textwidth}
\subfloat[\ding{202} \textbf{Scratch}. \label{fig:costeff-scratch}]{
  \begin{minipage}[b]{0.352\textwidth}
    \includegraphics[width=\textwidth]{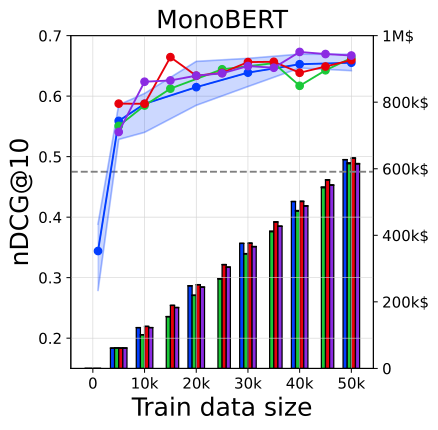}
  \end{minipage}
  \hfill
  \begin{minipage}[b]{0.33\textwidth}
    \includegraphics[width=\textwidth]{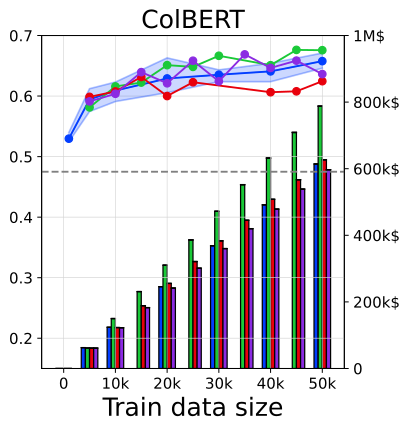}
  \end{minipage}
  \hfill
  \begin{minipage}[b]{0.33\textwidth}
    \includegraphics[width=\textwidth]{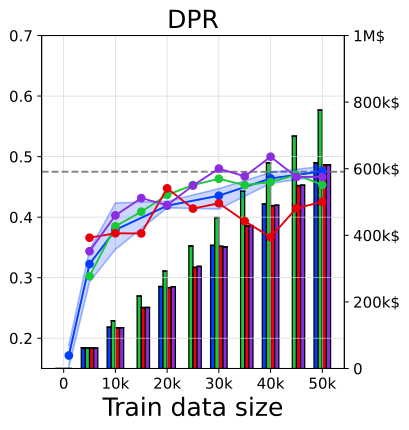}
  \end{minipage}}
\end{minipage}
\hfill 
\begin{minipage}{1.2\textwidth}
  \subfloat[\ding{203} \textbf{Re-Train}. \label{fig:costeff-retrain}]{
  \begin{minipage}[b]{0.29\textwidth}
    \includegraphics[width=\textwidth]{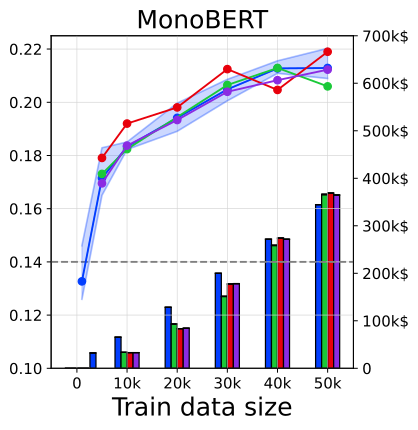}
  \end{minipage}
  \hfill
  \begin{minipage}[b]{0.29\textwidth}
    \includegraphics[width=\textwidth]{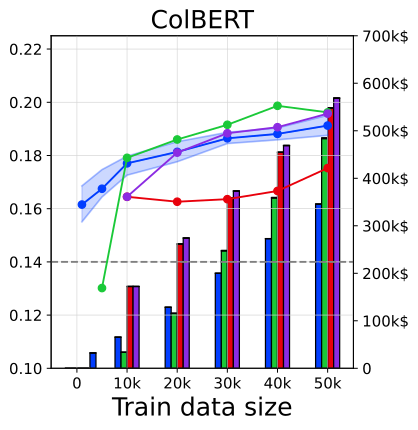}
  \end{minipage}
  \hfill
  \begin{minipage}[b]{0.43\textwidth}
    \includegraphics[width=\textwidth]{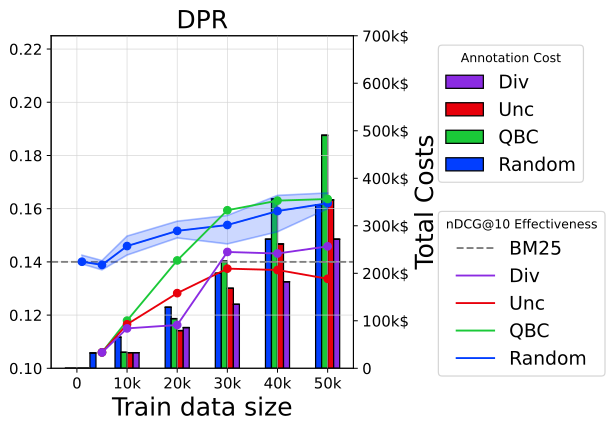}
  \end{minipage}
  }
\end{minipage}
}
\caption{nDCG@10 (lines, left y-axis) and stacked annotation and computational cost (bars, right y-axis) for different train data sizes on TREC DL 2020 (\ding{202} \textbf{Scratch}, Figure \ref{fig:costeff-scratch}) and on TripClick Head DCTR (\ding{203} \textbf{Re-Train}, Figure \ref{fig:costeff-retrain}). For Random (4 runs) the blue line denotes mean, shaded area denotes the range between min and max effectiveness. Good results would be expected to be between mean and max of Random, bad results between mean and min. For stacked cost, only annotation cost is visible since it greatly exceed computational costs.}
\label{fig:costeff}
\vspace{-10pt}
\end{figure*}

\begin{figure*}[t]
\centering
\captionsetup[subfloat]{farskip=0.5pt,captionskip=0.5pt}
\resizebox{1\linewidth}{!}{
\begin{minipage}{0.85\textwidth}
\subfloat[\ding{202} \textbf{Scratch}. \label{fig:ann-scratch}]{
  \begin{minipage}[b]{0.345\textwidth}
    \includegraphics[width=\textwidth]{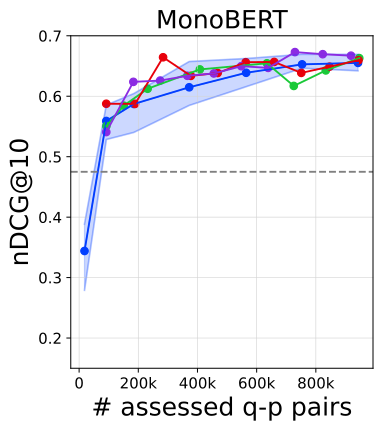}
  \end{minipage}
  \hfill
  \begin{minipage}[b]{0.33\textwidth}
    \includegraphics[width=\textwidth]{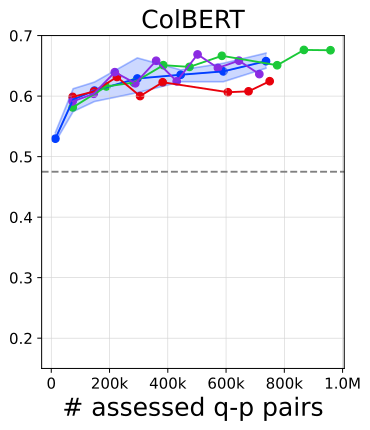}
  \end{minipage}
  \hfill
  \begin{minipage}[b]{0.315\textwidth}
    \includegraphics[width=\textwidth]{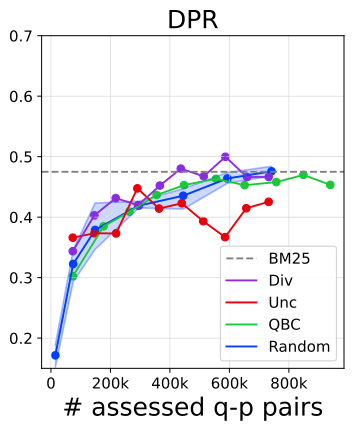}
  \end{minipage}
  }
\end{minipage}
  \hfill
\begin{minipage}{0.82\textwidth}
  \subfloat[\ding{203} \textbf{Re-Train}. \label{fig:ann-retrain}]{
  \begin{minipage}[b]{0.33\textwidth}
    \includegraphics[width=\textwidth]{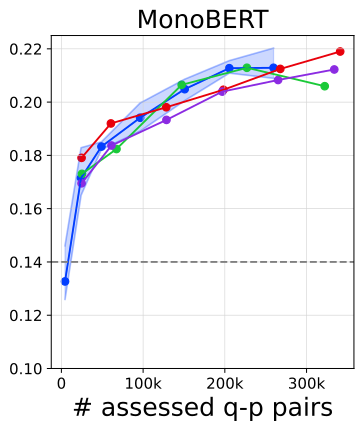}
  \end{minipage}
  \hfill
  \begin{minipage}[b]{0.33\textwidth}
    \includegraphics[width=\textwidth]{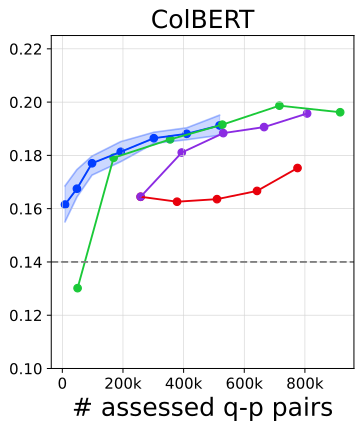}
  \end{minipage}
  \hfill
  \begin{minipage}[b]{0.33\textwidth}
    \includegraphics[width=\textwidth]{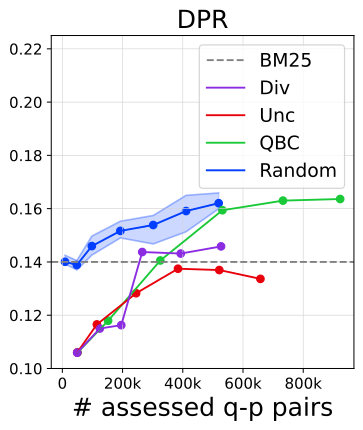}
  \end{minipage}
  }
\end{minipage}
}
\caption{nDCG@10 vs. number of assessed query-passage pairs on TREC DL 2020 (\ding{202} \textbf{Scratch}, Figure \ref{fig:ann-scratch}) and on TripClick Head DCTR (\ding{203} \textbf{Re-Train}, Figure \ref{fig:ann-retrain}). Number of assessments per sample is measured with rank of highest relevant passage during selection. For the Random baseline the blue line denotes mean, blue shaded area denotes the range between max and min effectiveness versus mean of the number of assessments. Selection strategies are not consistently more effective considering the number of assessments to annotate the training samples.}
\label{fig:ann}
\vspace{-12pt}
\end{figure*}

The effectiveness results are more consistent across methods and training data sizes in scenario \ding{203} \textbf{Re-Train}. Random outperforms all AL selection strategies when using DPR.
The QBC strategy reaches slightly higher effectiveness than random selection when ColBERT is used; however, none of the improvements are significant despite the large number of test queries in the TripClick Head and Torso test sets. No statistical significance is found even when the worst random selection run is considered in place of the mean of the random runs.


In summary, we found that for the task of fine-tuning PLM rankers, there is no single active learning selection strategy that consistently and significantly delivers higher effectiveness compared to a random selection of the training data. This is a surprising and interesting result. Active learning has been shown to be effective in natural language tasks \cite{lewis1994uncertaintysigir}, also for methods that rely on PLM models \cite{ein-dor-etal-2020-active}: yet, popular AL methods do not work in the context of PLM rankers. However, RQ1 shows that there are subsets of the training data that when used for fine-tuning PLM rankers deliver sensibly higher effectiveness than others -- but AL methods are unable to identify those high-yield training samples.


\vspace{-4pt}
\subsection{RQ3: Budget-aware Evaluation}

Since the goal of actively selecting training data is to minimize the annotation cost, we investigate the active selection strategies in the context of constraint budgets. For this, we use the budget-aware evaluation of Section~\ref{sec:budgetaware}, which accounts for the number of assessments needed to annotate the training data as well as the computational cost of the training and selection.

We visualize the effectiveness and associated costs at different training set sizes for the AL strategies for the three PLM rankers in Figure \ref{fig:costeff-scratch} for \ding{202} \textbf{Scratch} on TREC DL 2020 and in Figure \ref{fig:costeff-retrain} for \ding{203} \textbf{Re-Train} on TripClick Head DCTR. The lines and the left y-axis refer to the rankers' effectiveness, measured as nDCG@10. The bars and the right y-axis refer to the total cost computed with the budget-aware evaluation. The bars are stacked (the annotation and computational cost), but since with our cost settings the annotation cost greatly exceeds the computational cost, the bars for the GPU and CPU costs are not visible.
In all figures the blue line denotes the effectiveness of Random, with the blue shade representing the range measured between the worst and best random selection runs (recall that random selection was ran four times, and Random is the mean effectiveness of these runs).

A first observation is that the main cost factor is the annotation cost, and hence the number of assessments needed to create the training data, which largely overrules the computational cost. Because of this, in Figures~\ref{fig:ann-scratch} (\ding{202} \textbf{Scratch}) and~\ref{fig:ann-retrain} (\ding{203} \textbf{Re-Train})
we further visualise the effectiveness of the AL strategies relative to the number of assessments needed to reach that effectiveness.



Next, we analyse the results for \ding{202} \textbf{Scratch} (Figures~\ref{fig:costeff-scratch} and~\ref{fig:ann-scratch}).
For MonoBERT, the active selection strategies often provide higher effectiveness than Random when more than 10k samples are available -- these effectiveness gains are however not significant. Nonetheless, QBC and diversity require a lower budget than the Random baseline with savings of up to 15k\$ when 50k query-document samples are collected.  We note that the uncertainty-based strategy provides similar effectiveness to Random (especially from 20k samples), at no cost-savings.


For ColBERT, QBC consistently provides higher effectiveness than Random, however at a much higher cost. For example, when 50k training samples are selected, using QBC costs nearly \$200k more than Random, requiring annotations for roughly 200k more query-document pairs. In fact, when approximately the same budget/number of annotations are used, QBC and Random obtain the same effectiveness (in Figure~\ref{fig:ann-scratch} compare the last point of Random with the third last point of QBC). Aside from QBC, all other active selection strategies deliver similar or lower effectiveness of Random, for the same or higher cost.


For DPR, the uncertain-based strategy consistently delivers inferior effectiveness than the baseline. QBC and diversity-based selection do provide effectiveness gains when the training data is in the range 10k to 40-45k samples. For QBC, however, these gains come at a large budget expense: for 30k the QBC selection requires 90k\$ more annotation budget than Random. The diversity-based strategy instead does deliver some costs-savings compared to Random. For example, for Random to reach the same effectiveness of BM25, about 600k annotations are needed, while diversity delivers the same level of effectiveness with only 420k annotations. However, we note using more annotations with diversity-based sampling does not necessarily translate in a more effective model: going from 600k to about 750k annotations deteriorates the search effectiveness of the ranker.


Looking across PLM rankers, we observe that while the annotation costs across selection strategies are relatively similar for MonoBERT, they are higher for QBC than all other strategies when ColBERT and DPR are considered.

Overall, the selection strategies show relatively unstable effectiveness, the effectiveness can even decrease when training data increases. This is particularly the case for uncertainty selection for DPR: for example, its effectiveness decreases by from 0.45 to 0.37 when the amount of training data doubles from 20k to 40k.

We next analyse the results for scenario \ding{203} \textbf{Re-Train}.
While some selection strategies provide gains over random selection, these gains largely depend on which PLM is used and the training size (Figure \ref{fig:costeff-retrain}). Nevertheless, despite the specific gains in effectiveness, all active selection strategies require more assessments, and thus a higher budget, to reach the same level of effectiveness obtained when using random selection (Figure \ref{fig:ann-retrain}).


For MonoBERT, uncertainty selection exhibits (non-significant) improvements when training data is less than 30k. In fact, for small amounts of training data, uncertainty sampling does provide some cost savings: for example MonoBERT with uncertainty sampling needs about 65,000 query-passage pairs assessments to obtain the same effectiveness obtained with random selection with $\approx100k$ assessed pairs. However, this effect is lost when the training data size increases further, with the budget required by uncertainty sampling becoming similar (or more in some instances of random selection) to that of Random to obtain the same level of effectiveness. All other active selection strategies, when used with MonoBERT, deliver either lower effectiveness than Random, or higher costs. This is the case particularly for QBC. In fact, although there is one setting in which QBC delivers major cost savings to reach the same effectiveness of the random baseline (QBC achieves nDCG@10 higher than 0.2 using a sensibly lower amount of annotated query-passage pairs), cost savings are not consistent across all training data sizes and larger sizes correspond to a higher number of assessments required compared to Random.


For ColBERT, QBC and diversity selection outperform the baseline from training data sizes of 20k onward. This however comes with a considerable increase of query-passage pairs to be assessed and thus of annotation cost. For example, with a training subset of 30k, random selection costs about \$200,000 while diversity selection costs nearly double that -- but the increase in search effectiveness is marginal. It is interesting to compare these results with that obtained for scenario \ding{202} \textbf{Scratch}. While in both scenarios uncertainty selection shows effectiveness losses when more training data is added, and QBC is associated with higher costs, diversity selection performs differently: it provides similar effectiveness for a similar cost in \ding{202} \textbf{Scratch}, and a marginal effectiveness improvement for a largely higher cost in \ding{203} \textbf{Re-Train}.



For DPR, all selection strategies underperform random selection, with the exception of QBC that provides marginal improvements when the training subset is larger than 30k, but this at the expense of a higher budget. The budget-aware evaluation, in fact, shows that all selection strategies require more query-passage pairs to be assessed (higher cost) than the random selection baseline to reach the same search effectiveness (and some strategies cannot even achieve that effectiveness). An example is QBC that requires 730,000 assessments to reach the same effectiveness obtained by Random with just $\approx250k$ assessments.


In summary, in answer to RQ3, we found that the use of the investigated active selection strategies does not deliver consistent budget savings. In our experiments, the budget is largely dominated by the assessment cost and all active selection strategies tend to require a higher amount of query-passage pairs to be annotated than random selection. Even in contexts where assessment is very cheap, active selection would not provide budget savings because more assessments are required for active selection than for random selection. We do note that there are cases where specific active selection strategies provide similar search effectiveness than random selection at a reduced cost. However, these cases occur for specific choices of selection strategy, PLM ranker and training subset size and thus are unlikely to generalise in practice.


\vspace{-10pt}
\section{Conclusion} \label{conclusions}

We investigated fine-tuning PLM rankers under limited data and budget.
For this, we adapted several active selection strategies, representing different key approaches in active learning that have been shown effective in many natural language processing tasks. Surprisingly, we found that for the task of fine-tuning PLM rankers no AL strategy consistently and significantly outperformed random selection of training data. However we found that there are subsets of the training data which lead to significantly higher effectiveness than others, thus we see it as an important open challenge to be able to automatically identify those training samples.
Similarly, our budget-aware evaluation showed that the investigated AL strategies do not deliver consistent \textit{budget savings }since they require a higher amount of assessments than random selection. 

One limitation of our study is that the estimation of annotation costs relies on sparse annotations of the training set. Potentially, the required number of assessments could be lower, since another relevant passage -- that is not marked as relevant in the data -- could be found earlier in the ranked list. We argue, however, that this should affect all selection strategies and does not benefit one strategy particularly. 

Another limitation is the way uncertainty was computed in our experiments. 
Uncertainty estimation in Information Retrieval is a fundamental but largely unexplored problem~\cite{turtle1997uncertainty,crestani1998information,collins2007estimation}, especially for rankers based on PLMs~\cite{lesota2021modern,cohen2021not}. Attempts have been made to exploit uncertainty in relevance estimation for traditional statistical models such as language models and BM25~\cite{zhu2009risky,wang2009portfolio}, but in these works the actual estimation of uncertainty is based on assumptions and heuristics such as to be related to similarities or covariance between term occurrences~\cite{zhu2009risky,wang2009portfolio,zuccon2009quantum}, to follow the Dirichlet distribution~\cite{wang2009portfolio}, or to be computed based on score distributions obtained through query term resampling~\cite{collins2007estimation}. Recent attempts have been made to model uncertainty for neural rankers, for example Transformer Pointer Generator Network (T-PGN) model~\cite{lesota2021modern}, or Cohen et al.'s~\cite{cohen2021not} efficient uncertainty and calibration modelling strategies based on Monte-Carlo drop-out~\cite{gal2016dropoutal}, but these are not readily applicable to the PLM ranker architectures we consider. In future work we plan to adapt and investigate these uncertainty estimations. 

Finally we also highlight that we only considered common baseline active learning methods. More sophisticated AL methods exist~\cite{ash2019deep,yuan2020cold,margatina2021active}, including alternating between selection types like in AcTune, which alternates active
learning and self-training~\cite{yu2021actune}, and  Augmented SBERT which alternates random selection and kernel density estimation based selection~\cite{thakur2021augmented}.
However, each of these approaches present specific challenges to be adapted to ranking. We also were interested to understand the promise AL has for PLM-based rankers, and provide a framework, inclusive of evaluation methodologies and baselines, in which these more advanced methods could be studied.


\bibliographystyle{ACM-Reference-Format}
\bibliography{sigir2023-al-for-rankers}

\end{document}